\documentclass[12pt]{article}
\begin{document}
\baselineskip .3in

\begin{center}
{\large{\bf Quark binding potential and QGP}}

 \vskip.2in

  SHUKLA PAL$^{1}$, APARAJITA BHATTACHARYA$^{2}$, BALLARI CHAKRABARTI$^{3}$,
  RISMITA GHOSH$^{4}$

\vskip .1in

{\small{$^{1,2,4}$ Department of Physics, Jadavpur University
Kolkata 700032, India.\\
$^{3}$Department of Physics,Jogamaya Devi College,Kolkata, India.\\
e.mail: $^{2} pampa@phys.jdvu.ac.in$}}

\end{center}

\vskip .2in
{\small{ The effect of quark-antiquark potential on the dissociation energy and critical screening length of heavy meson like $b\overline{b}$ and $c\overline{c}$ have been investigated when the respective meson is in quark Gluon Plasma(QGP). The different types of interquark potential have been used which get screened in the QGP medium. The dissociation energy and critical screening length have been studied for both ground and excited states. It has been observed that the form of interquark potential has substantial effect on the critical screening length when the meson is in QGP. A comparison with other theoretical studies are made.}

\vskip .5in

\hskip 2in {\bf 1. Introduction}

\vskip .2in

 It is well known that the strongly interacting matter with a very high density undergoes a transition to a state of deconfined  quarks and
 gluons. Deconfinement takes place if the color screening dissolves the binding potential between quark and quark or quark and antiquark.
 The $J/\psi$ or the $\Upsilon$ have much smaller radii than the radii of usual mesons and nucleons, due to which the bound state remains
 unaffected in QGP unless and until the temperature or the density becomes so high that the binding of bound state gets broken. The suppression
 of $J/\psi$ is the one of the signals for quark deconfinement [1]. To investigate quark plasma formation experimentally, it is essential to depend on  color screening and deconfinement. In QGP medium, the string tension between a charm c and a charm antiquark $\bar{c}$ disappears
 and quarks and gluons are deconfined. Only the Coulomb type of color interaction exists between the c and the $\bar{c}$. After the deconfinement
 of $J/\psi$, it is impossible to create it by hadronization of plasma. Matsui [2] has discussed the heavy quark $J/\psi$ suppression as signature of quark gluon plasma formation. Considering the various attributes of deconfinement test, Satz [3] has concluded that the $J/\psi$ peak in the spectrum of lepton pairs which are emitted during nuclear collisions can give important and required information. At ultra relativistic energies, dynamical analysis of matter during nuclear collision has been done by Matsui [4]. He also studied $J/\psi$ suppression as a signature of QGP formation. Ruuskanen and Satz [5] have studied the dependence of longitudinal momentum of $J/\psi$ suppression and observed its effects on nuclear collision experiments. Karsch and Petronzio  [6] have used the concept of QGP to analyze the nuclear size which depends on  $J/\psi$ suppression in heavy ion collisions and  studied the transfer energy related to this heavy ion collision. Karsch et al [7] have observed the dependence of the dissociation energies, the binding radii and the masses of heavy quark resonances on the color screening length $r_{D}$ of the medium and concluded that no binding exists below $r_{D}$. Hence in high energy heavy ion collisions, the suppression of $J/\psi$ production may be considered as a symbol for the presence of quark gluon plasma. Bo.Liu and Yu-Bing Dong [8] have studied the binding and dissolution for the $c\bar{c}$ and $b\bar{b}$ bound states using different quark
 potentials in a non-relativistic approximation. They have estimated the critical value of the screening length using Debye screening effect. They have also estimated the critical temperature $T_{c}$
 of medium. Stubbins [9] has used the generalized variational method to investigate energies for the Yukawa and Hulthen potentials. A number of works have been done on the heavy meson dissociation in the recent time. Park[10] has investigated the heavy meson dissociation in light quark medium and discussed the mechanism of dissociation. He has observed that dissociation length decreases with increase of chemical potential in QGP whereas in hadronic phase they behave in opposite way. Braga et al.,[11] have studied thermal behavior of $c\overline{c}$ and $b\overline{b}$  by using holographic model and studied the effect of magnetic field on the thermal spectrum of heavy meson whereas Blaizot et al.,[12] investigated the heavy quark  formation, dissociation in QGP in the framework of Langevin  Equation. They have pointed out that formation of bound states occurs if enough heavy quarks are present iin the system whereas dissociation occurs due to screening of the potential in the plasma. Gau et al.,[13] studied charmonium $c\overline{c}$ wave function at finite temperature using relativistic Schrodinger equation for spin singlet and triplet. They found relativistic correction to the $J/\psi$ dissociation temperature in QGP is between 7$\%$ to 13$\%$. They have studied wavefunction S,D states. They have obtained that with temperature both S and D wavefunction expands. Adil et al.,[14] have investigated medium induced dissociation probability of heavy meson and found that it is sensitive to the opacity of the quark gluon plasma and time dependance of its formation and evolution. A comprehensive review of QCD, QGP and Heavy quark meson suppression and production is done by Kisslinger[15].

 In the present work we have investigated the effect of $q\overline{q}$ potential on dissociation energy of $b\overline{b}$ and $c\overline{c}$ mesons in QGP. It is well known that a number of phenomenological potentials have usually been used to describe the inter quark potential. The respective potentials get screened in QGP due to interaction with different particles. It is very important to study how the form of the inter quark potential affects the dissociation energy and subsequently affect the critical screening length of the heavy mesons while they are in QGP medium. We use a variational method to study the system and a trial wave function has been used. The dissociation energy and critical screening lengths have been estimated for different form of inter quark potential. A study with the variation of variational parameter of the trial wavefunction have also been made. The critical screening length at which dissociation energy vanishes have been extracted for $b\overline{b}$ and $c\overline{c}$ mesons for their ground and excited states.

\vskip .1in

\hskip 2in {\bf 2. Methodology}

\vskip .2in

  The Hamiltonian for a $q\bar{q}$ system in non-relativistic approach can be represented as:
\begin{equation}
H(r,T) = \frac{\overrightarrow{p}^{2}}{2m_{re}}+V_{q\bar{q}(r,T)}
\end{equation}
where $m_{re}$ is the reduced mass of the $q\bar{q}$ system,$V_{q\bar{q}}$ $(r,T)$ is the inter quark binding potential. The exact nature of the binding potential of q$\bar{q}$ system is not known but a number of phenomenological potentials are suggested which are very successful in describing the binding energy of the system. We have considered four different quark-antiquark potentials like Cornell potential, Rosener potential, Harmonic potential and a combination of potential for our study. The expression for the potentials run as:
\vskip .1in
(a) Cornell potential [7]

\begin{equation}
V(r,0) = -\frac{\alpha_{s}}{r}+K^{'}r
\end{equation}

where $K^{'}$ is the coefficient for confinement and is taken to be 0.92 $GeV^{2}$ [16] , $\alpha_{s}$ is the parameter proportional to the strong coupling constant which is 0.471 [7] and r is radius of
heavy mesons.
\vskip .25in
(b)Rosner potential [17]

\begin{equation}
V(r,0) = -\frac{A(r^{-\alpha}-1)}{\alpha}+B
\end{equation}
with ${\alpha}$ = 0.12, A = 0.801 GeV and B = -0.772 GeV

\vskip .25in
(c) Harmonic potential [18]
\vskip .25in
\begin{equation}
V(r,0) = \delta r^{2}
\end{equation}

with $\delta$ = 0.08 $GeV^{3}$

\vskip .25in
(d) A combination of harmonic, linear and Coulomb potential [19,20]:
\vskip .25in
\begin{equation}
V(r,0) = ar^{2}+br-c/r
\end{equation}

with a = 0.142 $GeV^{3}$, b = 0.465 $GeV^{2}$ and c = 0.471

\vskip .1in

In QGP environment of quarks and gluons, the interquark potential gets modified due to the color screening so that the $q\overline{q}$ aforesaid potentials can be represented as:

\begin{equation}
V(r,T) =-\frac{Ze^{-\lambda r}}{r}[-\frac{\alpha_{s}}{r}+K^{'}r]
\end{equation}

\begin{equation}
V(r,T) =-\frac{Ze^{-\lambda r}}{r}[-\frac{A(r^{-\alpha}-1)}{\alpha}+B]
\end{equation}

\begin{equation}
V(r,T) =-\frac{Ze^{-\lambda r}}{r}[\delta r^{2}]
\end{equation}

\begin{equation}
V(r,T) =-\frac{Ze^{-\lambda r}}{r}[ ar^{2}+br-c/r]
\end{equation}

 where Z is a constant and is equal to 1 $GeV^{-1}$ and $\lambda$ is a temperature dependent screening parameter. We have
 parameterized the temperature dependence of $\lambda$ as $\lambda(T)=\lambda(0)[1-T/T_{C}]^{-0.2}$ [21]. The value of $\lambda(0)$ has given input as 0.2 GeV from [7]. Solution of equation (1) with the potential in (2),(3), (4) and (5) will lead a temperature dependent binding energy. To get an expression for binding energy we have considered a trial wave function from the work of Stubbins [9] with a spherical component which runs as:

 \begin{equation}
\Psi_{k} = B_{k}r^{k}e^{-(\beta /2)}Y_{l,m}(\theta,\phi)
\end{equation}
 where k = 0,1,2,..... and l = 0,1,... $B_{k}$ is normalization constant and can be represented as:

\begin{equation}
 B_{k} =[\frac{\beta^{2k+3}}{(2k+2)!}]^{1/2}
\end{equation}
where $\beta$ is variational parameter [9]. Schordinger equation with related eigenvalue can be represented as [7]:
\begin{equation}
[H(r,\lambda(T))-E_{n,l}(\lambda(T))]\Psi_{K}(r,\lambda(T)) =0
\end{equation}
where n is principal quantum number and l is orbital quantum number such that $l\leq(n-1)$. Dependence of $\lambda$ on temperature leads to an temperature dependent eigen energy.
\vskip .1in

The dissolution energy is the quantity which accounts the vanishing of bound states. At a fixed value of $\lambda(T)$ the dissolution energy of the bound state can be defined as [7],
\begin{equation}
E_{dis}^{n,l}(\lambda(T)) = V_{q\bar{q}}(r\rightarrow\infty, \lambda(T)) + E_{n l}(\lambda(T))
\end{equation}

At $r\rightarrow\infty$ the equation (13) reduces to:
\begin{equation}
E_{dis}^{n,l}(\lambda(T)) =  E_{n l}(\lambda(T))
\end{equation}

The value of dissolution energy is positive for bound states and is negative for continuum and leads to the condition:

\begin{equation}
E_{dis}^{n,l}(\lambda_{C}(T)) =  0
\end{equation}

Equation (15) gives the critical value of $\lambda(T)$, beyond which for given quantum number there are no bound states. We have considered first three radial excitation corresponding to $J/\psi$ and $\Upsilon$ for n=1, l=0, $\psi^{'}$ and $\Upsilon^{'}$ for n=2, l=0 and $\psi^{''}$ and $\Upsilon^{''}$ for n=3, l=0 and  $\chi_{c}$ and $\chi_{b}$ for n=2, l=1. We have derived the binding energies of above mentioned states and the expressions for binding energies of ground states, first and second excited states for four different potentials are obtained as:
\vskip.2in

(i)  Cornell Potential
\vskip.1in
(a) 1s-State
\begin{equation}
E_{BE} = -2\pi\beta^{3}[\frac{2K^{'}}{(\beta+\lambda)^{3}}-\frac{\alpha_{s}}{(\beta+\lambda)}]+\frac{\pi\beta^{2}}{2m_{r}}
\end{equation}
\vskip.1in
(b) 2s-state
\begin{equation}
E_{BE} = -\frac{\pi\beta^{5}}{3}[\frac{12K^{'}}{(\beta+\lambda)^{5}}-\frac{\alpha_{s}}{(\beta+\lambda)^{3}}]+\frac{\pi\beta^{2}}{6m_{r}}
\end{equation}
\vskip.1in
(c) 3s-state
\begin{equation}
E_{BE} = -\frac{2\pi\beta^{7}}{15}[\frac{30K^{'}}{(\beta+\lambda)^{7}}-\frac{\alpha_{s}}{(\beta+\lambda)^{5}}]+\frac{\pi\beta^{2}}{10m_{r}}
\end{equation}
\vskip.1in

(ii)Rosner Potential
\vskip.1in
(a)1s-State
\begin{equation}
E_{BE} = -2\pi\beta^{3}[\frac{(A/\alpha) + C}{(\beta+\lambda)^{2}}]+\frac{\pi\beta^{2}}{2m_{r}}
\end{equation}
\vskip.1in
(b) 2s-State
\begin{equation}
E_{BE} = -\frac{\pi\beta^{5}}{3}[\frac{(A/\alpha)+C}{(\beta+\lambda)^{3}}]+\frac{\pi\beta^{2}}{6m_{r}}
\end{equation}
\vskip.1in
(c) 3s-State
\begin{equation}
E_{BE} = -\frac{2\pi\beta^{7}}{3}[\frac{(A/\alpha)+C}{(\beta+\lambda)^{5}}]+\frac{\pi\beta^{2}}{10m_{r}}
\end{equation}
\vskip.1in

(iii)Harmonic Potential

\vskip.1in
(a)1s-State
\begin{equation}
E_{BE} = -[\frac{12\pi\beta^{3}a}{(\beta+\lambda)^{4}}]+\frac{\pi\beta^{2}}{2m_{r}}
\end{equation}
\vskip.1in
(b) 2s-State
\begin{equation}
E_{BE} = -\frac{20\pi\beta^{5}a}{(\beta+\lambda)^{6}}+\frac{\pi\beta^{2}}{6m_{r}}
\end{equation}
\vskip.1in
(c) 3s-State
\begin{equation}
E_{BE} =  -\frac{28\pi\beta^{7}a}{(\beta+\lambda)^{8}}+\frac{\pi\beta^{2}}{10m_{r}}
\end{equation}
\vskip.1in
and (iv) A combination of harmonic, linear and Coulomb potential

\vskip.1in
(a)1s-State
\begin{equation}
E_{BE} = -\frac{2\pi\beta^{3}}{(\beta+\lambda)}[\frac{6a}{(\beta+\lambda)^{3}}+\frac{2b}{(\beta+\lambda)^{2}}-c]+\frac{\pi\beta^{2}}{2m_{r}}
\end{equation}
\vskip.1in
(b) 2s-State
\begin{equation}
E_{BE} = -\pi\beta^{5}[\frac{20a}{(\beta+\lambda)^{6}}\frac{4b}{(\beta+\lambda)^{5}}-\frac{c}{3(\beta+\lambda)^{3}}]+\frac{\pi\beta^{2}}{6m_{r}}
\end{equation}
\vskip.1in
(c) 3s-State
\begin{equation}
E_{BE} = -2\pi\beta^{7}[\frac{14a}{(\beta+\lambda)^{8}}+\frac{2b}{(\beta+\lambda)^{7}}-\frac{c}{15(\beta+\lambda)^{5}}]+\frac{\pi\beta^{2}}{10m_{r}}
\end{equation}
\vskip.1in

We have estimated the critical $\lambda(\lambda_{C}(T))$  and critical radius using the equation (15) with different values of variational parameter $\beta$ and the results are furnished in Table No.1, 2, 3 and 4.
\vskip 2in
\hskip 5in {\bf Table I}
\vskip .1in
 Temperature dependent critical screening lengths $\lambda_{c}$ in GeV with different values of variational parameter $(\beta)$ in GeV using Cornell Potential.
 \vskip .1in

\begin{tabular}{|r |r |r |r |r |r |r|}
  \hline
 \hline
  States &\hskip .1in&\hskip .1in$\beta$=0.1 &\hskip .1in $\beta$=0.2 &\hskip .1in$\beta$=0.3&\hskip .1in$\beta$=0.4&\hskip .1in$\beta$=0.5\\
         &\hskip .1in&\hskip .1in GeV        &\hskip .1in GeV          &\hskip .1in GeV      &\hskip .1in GeV       &\hskip  .1in GeV\\

  \hline
  $J/\psi$(n=1, l=0) & $\lambda_{c}$(GeV)&$ 0.6391$&                  $0.7048$&                         $0.7127$&                $0.6924$&         $0.6577$\\
           & $r_{C}$(fm)          & $0.3139$&               $0.2873$&                         $0.2806$&                $0.2888$&         $0.3041$\\
  $\psi^{'}$(n=2, l=0)& $\lambda_{c}$(GeV)&$ 0.3274$&                 $0.4437$&                         $0.5189$&                $0.5706$&         $0.5975$\\
            & $r_{C}$(fm)          &$0.6108$&               $0.4509$&                         $0.3854$&                $0.3503$&         $0.3347$\\
  $\psi^{''}$(n=3, l=0)& $\lambda_{c}$(GeV)&$ 0.2042$&                $0.2963$&                         $0.3632$&                $0.4140$&         $0.4466$  \\
            & $r_{C}$(fm)         &$0.979$&                 $0.6749$&                         $0.5506$&                $0.3545$&         $0.4478$\\
   $\chi_{c}$(n=2, l=1)& $\lambda_{c}$(GeV) &$0.3641$&                $0.4395$&                         $0.5153$&                $0.5682$&         $0.5771$  \\
            &$r_{C}$(fm)        &$0.5493$&                  $0.4550$&                         $0.3881$&                $0.3519$&         $0.3465$\\
   $\Upsilon$(n=1, l=0)& $\lambda_{c}$(GeV)&$ 0.9341$&                $1.0353$&                         $1.0506$&                $1.0315$&         $0.9938$\\
            &$r_{C}$(fm)        &$0.2141$&                  $0.1932$&                         $0.1903$&                $0.1939$&         $0.2012$\\
  $\Upsilon^{'}$(n=2, l=0)& $\lambda_{c}$(GeV)&$0.4331$&              $0.6197$&                         $0.7342$&                $0.8244$&         $0.9146$\\
               &$r_{C}$(fm)         &$0.4618$&              $0.3227$&                         $0.2724$&                $0.2426$&         $0.2187$\\

   $\Upsilon^{''}$(n=3, l=0)& $\lambda_{c}$(GeV)&$0.2639$&            $0.3910$&                         $0.4860$&                $0.5641$&         $0.6197$\\
                    &$r_{C}$(fm)      &$0.8442$&            $0.5115$&                         $0.4115$&                $0.4831$&         $0.3227$\\
   $\chi_{b}$ (n=2, l=1)&  $\lambda_{c}$(GeV)    &$0.5023$&            $0.5893$&                         $0.7160$&                $0.8132$&         $0.8551$\\
            & $r_{C}$(fm)             &$0.3981$&            $0.3394$&                         $0.2793$&                $0.2459$&         $0.2339$\\

\hline

\end{tabular}

\vskip 1in
\hskip 5in {\bf Table II}
\vskip .1in

Temperature dependent critical  screening lengths $\lambda_{c}$ in GeV with different values of variational parameter $(\beta)$ in GeV using Rosner Potential.
 \vskip .1in

\begin{tabular}{|r |r |r |r |r |r |r|}
  \hline
\hline
  States &\hskip .1in&\hskip .1in$\beta$=0.1 &\hskip .1in $\beta$=0.2 &\hskip .1in$\beta$=0.3&\hskip .1in$\beta$=0.4&\hskip .1in$\beta$=0.5\\
         &\hskip .1in&\hskip .1in GeV        &\hskip .1in GeV          &\hskip .1in GeV      &\hskip .1in GeV       &\hskip  .1in GeV\\

  \hline
  $J/\psi$(n=1, l=0) & $\lambda_{c}$(GeV) &$1.12$&                  $1.536$&                         $1.827$&                $2.057$&         $2.248$\\
           & $r_{C}$(fm)        &$0.178$&                 $0.130$&                         $0.109$&                $0.097$&         $0.088$\\
  $\psi^{'}$(n=2, l=0)& $\lambda_{c}$(GeV)&$ 0.095$&                $0.191$&                         $0.287$&                $0.384$&         $0.480$\\
            & $r_{C}$(fm)       &$2.105$&                 $1.047$&                         $0.697$&                $0.521$&         $0.417$\\
  $\psi^{''}$(n=3, l=0)& $\lambda_{c}$(GeV)&$ 0.089$&               $0.180$&                         $0.271$&                $0.362$&         $0.453$  \\
            & $r_{C}$(fm)        &$2.247$&                $1.111$&                         $0.738$&                $0.552$&         $0.441$\\
   $\chi_{c}$(n=2, l=1)& $\lambda_{c}$(GeV)&$0.07$&                 $0.182$&                         $0.283$&                $0.379$&         $0.478$  \\
            &$r_{C}$(fm)         &$2.857$&                $1.099$&                         $0.707$&                $0.528$&         $0.418$\\
   $\Upsilon$(n=1, l=0)& $\lambda_{c}$(GeV)&$ 2.05$&                $2.93$&                          $3.55$&                 $4.05$&          $4.475$\\
            &$r_{C}$(fm)        &$0.0975$&                $0.0682$&                        $0.056$&                $0.049$&         $0.044$\\
  $\Upsilon^{'}$(n=2, l=0)& $\lambda_{c}$(GeV)&$0.182$&             $0.347$&                         $0.573$&                $0.762$&         $0.955$\\
               &$r_{C}$(fm)         &$1.099$&             $0.535$&                         $0.349$&                $0.262$&         $0.209$\\

   $\Upsilon^{''}$(n=3, l=0)& $\lambda_{c}$(GeV)&$0.137$&           $0.279$&                         $0.423$&                $0.566$&         $0.708$\\
                    &$r_{C}$(fm)      &$1.459$&           $0.717$&                         $0.473$&                $0.353$&         $0.282$\\
   $\chi_{b}$(n=2, l=0)&  $\lambda_{c}$(GeV)    &$0.145$&           $0.341$&                         $0.51$&                 $0.743$&         $0.936$\\
            & $r_{C}$(fm)             &$1.379$&           $0.586$&                         $0.392$&                $0.269$&         $0.213$\\

\hline

\end{tabular}

\vskip .5in
\hskip 5in {\bf Table III}
\vskip .1in
Temperature dependent critical screening lengths $\lambda_{c}$ in GeV with different values of variational parameter $(\beta)$ in GeV using Harmonic Potential.

 \vskip .1in

\begin{tabular}{|r |r |r |r |r |r |r|}
  \hline
\hline
  States &\hskip .1in&\hskip .1in$\beta$=0.1 &\hskip .1in $\beta$=0.2 &\hskip .1in$\beta$=0.3&\hskip .1in$\beta$=0.4&\hskip .1in$\beta$=0.5\\
         &\hskip .1in&\hskip .1in GeV        &\hskip .1in GeV          &\hskip .1in GeV      &\hskip .1in GeV       &\hskip  .1in GeV\\

  \hline
  $J/\psi$ (n=1, l=0)& $\lambda_{c}$(GeV)&$ 0.4893$&                  $0.5034$&                         $0.4789$&                $0.4371$&         $0.3846$\\
           & $r_{C}$(fm)          & $0.4087$&               $0.3973$&                         $0.4176$&                $0.4575$&         $0.52$\\
  $\psi^{'}$(n=2, l=0)& $\lambda_{c}$(GeV)&$ 0.3237$&                 $0.4041$&                         $0.4409$&                $0.4556$&         $0.4567$\\
            & $r_{C}$(fm)          &$0.6178$&               $0.4949$&                         $0.4536$&                $0.4389$&         $0.4379$\\
  $\psi^{''}$(n=3, l=0)& $\lambda_{c}$(GeV)&$ 0.2297$&                $0.3093$&                         $0.3567$&                $0.3863$&         $0.4041$  \\
            & $r_{C}$(fm)         &$0.8707$&                $0.6466$&                         $0.5607$&                $0.5177$&         $0.4949$\\
   $\chi_{c}$(n=2, l=1)& $\lambda_{c}$(GeV) &$0.2923$&                $0.4006$&                         $0.4343$&                $0.4518$&         $0.4539$  \\
            &$r_{C}$(fm)        &$0.6842$&                  $0.4992$&                         $0.4573$&                $0.4427$&         $0.4406$\\
   $\Upsilon$(n=1, l=0)& $\lambda_{c}$(GeV)&$ 0.6931$&                $0.7444$&                         $0.7476$&                $0.7266$&         $0.6911$\\
            &$r_{C}$(fm)        &$0.2885$&                  $0.2687$&                         $0.2675$&                $0.2752$&         $0.2894$\\
  $\Upsilon^{'}$(n=2, l=0)& $\lambda_{c}$(GeV)&$0.4118$&              $0.5335$&                         $0.6004$&                $0.6413$&         $0.6656$\\
               &$r_{C}$(fm)         &$0.4857$&              $0.3749$&                         $0.3331$&                $0.3118$&         $0.3005$\\

   $\Upsilon^{''}$(n=3, l=0)& $\lambda_{c}$(GeV)&$0.2759$&            $0.3899$&                         $0.4602$&                $0.5108$&         $0.5475$\\
                    &$r_{C}$(fm)      &$0.7249$&            $0.5129$&                         $0.4345$&                $0.3915$&         $0.3653$\\
   $\chi_{b}$(n=2, l=1)&  $\lambda_{c}$(GeV)    &$0.3511$&            $0.5023$&                         $0.5806$&                $0.6305$&         $0.6508$\\
            & $r_{C}$(fm)             &$0.5696$&            $0.3981$&                         $0.3445$&                $0.3172$&         $0.3073$\\

\hline

\end{tabular}

\vskip .5in
\hskip 5in {\bf Table IV}
\vskip .1in
Temperature dependent critical screening lengths $\lambda_{c}$  in GeV with different values of variational parameter $(\beta)$ in GeV using potential IV.
 \vskip .2in

\begin{tabular}{|r |r |r |r| r| r| r|}
  \hline
\hline
  States &\hskip .1in&\hskip .1in$\beta$=0.1 &\hskip .1in $\beta$=0.2 &\hskip .1in$\beta$=0.3&\hskip .1in$\beta$=0.4&\hskip .1in$\beta$=0.5\\
         &\hskip .1in&\hskip .1in GeV        &\hskip .1in GeV          &\hskip .1in GeV      &\hskip .1in GeV       &\hskip  .1in GeV\\

  \hline
  $J/\psi$(n=1, l=0) & $\lambda_{c}$(GeV)&$ 0.6656$&                  $0.7095$&                         $0.7014$&                $0.6679$&         $0.6204$\\
           & $r_{C}$(fm)          & $0.3005$&               $0.2819$&                         $0.2851$&                $0.2994$&         $0.3224$\\
  $\psi^{'}$(n=2, l=0)& $\lambda_{c}$(GeV)&$ 0.3910$&                 $0.5075$&                         $0.5759$&                $0.6194$&         $0.6464$\\
            & $r_{C}$(fm)          &$0.5115$&               $0.3941$&                         $0.3473$&                $0.3229$&         $0.3094$\\
  $\psi^{''}$(n=3, l=0)& $\lambda_{c}$(GeV)&$ 0.2599$&                $0.3626$&                         $0.4319$&                $0.4813$&         $0.5185$  \\
            & $r_{C}$(fm)         &$0.7695$&                $0.5516$&                         $0.4631$&                $0.4155$&         $0.3857$\\
   $\chi_{c}$(n=2, l=1)& $\lambda_{c}$(GeV) &$0.3612$&                $0.5023$&                         $0.5721$&                $0.6162$&         $0.6435$  \\
            &$r_{C}$(fm)        &$0.5537$&                  $0.3081$&                         $0.3496$&                $0.3245$&         $0.3108$\\
   $\Upsilon$(n=1, l=0)& $\lambda_{c}$(GeV)&$ 0.9183$&                $0.9822$&                         $0.9767$&                $0.9404$&         $0.8878$\\
            &$r_{C}$(fm)        &$0.2178$&                  $0.2036$&                         $0.2048$&                $0.2127$&         $0.2253$\\
  $\Upsilon^{'}$(n=2, l=0)& $\lambda_{c}$(GeV)&$0.4928$&              $0.6683$&                         $0.7784$&                $0.8551$&         $0.9112$\\
               &$r_{C}$(fm)         &$0.4058$&              $0.2997$&                         $0.2569$&                $0.2339$&         $0.2195$\\

   $\Upsilon^{''}$(n=3, l=0)& $\lambda_{c}$(GeV)&$0.3169$&            $0.4507$&                         $0.5510$&                $0.6271$&         $0.6877$\\
                    &$r_{C}$(fm)      &$0.6311$&            $0.4437$&                         $0.3629$&                $0.3189$&         $0.2908$\\
   $\chi_{b}$(n=2, l=1)&  $\lambda_{c}$(GeV)    &$0.4632$&            $0.6271$&                         $0.7677$&                $0.8325$&         $0.8973$\\
            & $r_{C}$(fm)             &$0.4318$&            $0.3189$&                         $0.2605$&                $0.2402$&         $0.2228$\\

\hline

\end{tabular}
\vskip .5in
Table I-IV displays our results of variation of screening length with variational parameter. Variational technique has provided an effective framework for spectroscopic studies of full hadron spectrum and a good candidate for investigation of strongly interacting system and gauge theories. The variational method has seen significance success in spectroscopic studies of hadronic system. A number of work have been done in QCD, lattice QCD applying the variational approach [22-24] and found to offer a more efficient method for the determination of nuclear matrix element [25]. Vega et al.,[26] have studied Cornell Potential using a trial wave function and super symmetric quantum mechanics. The parameters are changed applying successive transformation to obtain wave function at the origin. Ghalenvi et al.,[27] have studied baryon meson properties using trial wave function whereas Chot et al.,[28] have studied ground state masses with hyperfine interaction in QCD motivated effective Hamiltonian using a trial wave function. We have used a trial wave function to estimate the binding energy of mesons in QGP  and studied the variation of $\lambda_{c}$ with variational parameter.

\vskip .5in
\hskip 2in {\bf 3.Conclusions}
\vskip .2in
In the present work we have investigated the dissolution energy of heavy quarkonia $b\overline{b}$ and $c\overline{c}$ considering the effect of QGP medium in the interquark potential of the heavy mesons. We have used variational method to get the expression for energy and studied the critical parameter with the variation of the variational parameter of the trial wave function. We have also suggested an empirical form of temperature dependent screening parameter by the relation $\lambda(T)=\lambda(0)[1-T/T_{C}]^{-0.2}$ with the critical exponent 0.2 in our work. The study of critical phenomenon and corresponding scaling behavior of phase transition get a new impetus with the recent experimental development of studying low temperature physics. The quasi particle effective mass have been studied by parameterizing the behavior as $m^{*}=m(0)[{1-\frac{T}{T_{c}}}]^{\beta}$ [21]. Usually the critical point is reached by tuning the thermodynamic parameters. The renormalization group theory does not restrict the continuous variation of critical exponent which leads to the weak universality. Crystal behavior shows asymmetry in critical exponent $(\alpha)$ with $(\alpha)= -0.2 \pm0.3$. The critical exponent for $T<T_{c}$ for fluid varies from 0.1 to 0.2 [29]. The meanfield prediction of $(\alpha)= 0.5$ does not match with experimental value of 0.3 for fluid. The critical exponent is suggested to be 0.1 for $CO_{2}$ whereas for Xe the value is $\sim 0.2$ which do not violate the Rushbrooke or Griffith inequality. We have used critical exponent as treating the QGP as fluid and have studied the phase transition to estimate the critical screening length for heavy meson dissciation in QGP.   The variation of dissociation energy with variational parameter have been studied for ground states, first and second excited states of the heavy mesons. We have estimated the critical values of the screening parameters and critical radii for the states considering variational parameter $\beta$ varying from 0.1 GeV to 0.5 GeV. Liu et al [8] have studied the quark binding potential in QGP for various value of power of potential between quark and antiquark and studied $J/\psi$ suppression.  They have estimated the screening masses and Debye screening radii with temperature dependent different types of potentials and studied the values with different values of screening parameter. In the current work we have studied screening length and screening radii with the variation of variable parameter. Variational method is an useful tool to estimate the ground state energies and also excited states. It may be mentioned that variational method together with physically motivated trial wave function provide a powerful tool to study the systems under extreme condition and can be more robust in situation where it is difficult to determine a good unperturbed Hamiltonian. In the current work it has been observed that the value of critical screening lengths with variational parameter $\beta$ = 0.2 GeV estimated in present work with Cornell potential agree well with the estimation of Liu et al [8] with the value of screening parameter equal to 1.0 with model I. More comparative study will be done with this variational approach in our future work.

\vskip .2in
{\bf Acknowledgement}
\vskip .25in
The authors are thankful to the University Grants Commission, New Delhi, India, for financial support.
\vskip .25in

\hskip 2in{\bf References}

\vskip .1in

\noindent [1]  T. Matsui,  H. Satz, \emph{Phys. Lett. B} {\bf178}, 416 (1986).

\noindent [2]  T. Matsui,  \emph{Z. Phys. C} {\bf38}, 245 (1988).

\noindent [3]  H. Satz, \emph{Nucl. Phys. A} {\bf488}, 511c (1988).

\noindent [4]  T. Matsui, \emph{ Nucl. Phys. A } {\bf488},535c (1988).

\noindent [5]  P. V. Ruuskanen, H. Satz, \emph{Z. Phys. C} {\bf37}, 623 (1988).

\noindent [6]  F. Karsch, R. Petronzio, \emph{Phys. Lett. B} {\bf212}, 255 (1988).

\noindent [7]  F. Karsch,  M. T. Mehr, H. Satz  \emph{Z. Phys. C } {\bf37}, 617 (1988).

\noindent [8]  B. Liu, Y.B. Dong, \emph{Commun. Theor. Phys.} {\bf26}, 425 (1996).

\noindent [9]  C. Stubbins,  \emph{Phys. Rev. A} {\bf48}, 220 (1993).

\noindent [10] C. Park, \emph{Phys. Rev. D} {\bf81}, 045009 (2010).

\noindent [11] N. R. F. Braga, L. F. Ferreira, \emph{Phys. Lett. B}, {\bf783}, 186 (2018).

\noindent [12] J. P. Blaoit, D. D. Boni, P. Faccioli, G. Garberoglio,\emph{Nucl. Phys. A} {\bf946}, 49 (2016).

\noindent [13] X. Guo, S. Shi, P. Zhuang, \emph{Phys. Lett. B} {\bf718}, 143 (2012).

\noindent [14] A. Adil, I. Vitev, \emph{Phys. Lett. B} {\bf649}, 139 (2007).

\noindent [15] L. S. Kisslinger, \emph{Int. J. Mod. Phys. A} {\bf17300083} (2017).

\noindent [16] Seth, K. K. \emph{arXiv:0912.2776v1[hep-ex]} (2009).

\noindent [17]  A. K. Grant, J. L. Rosner,  E. Rynes, \emph{Phys. Rev. D} {\bf47},  1981 (1993).

\noindent [18]  M. R. Shojaei, A. A. Rajabi, \emph{Mod. Phys. Lett.} {\bf A23}, 3411 (2009).

\noindent [19]  R. Kumar, D. Kumar, F. Chand, \emph{Proceedings of DAE Symp. Nucl. Phys.} {\bf57}, 664 (2012).

\noindent [20]  M. R. Shojaei, H. T. Anbaran, \emph{Appl. Phys. Res.} {\bf2(1)}, 148 (2010).

\noindent [21]  A. Chandra, A. Bhattacharya, B. Chakrabarti, \emph{Jour. of Mod. Phys.} {\bf4} 945 (2013).

\noindent [22]  D. S. Roberts, W. Kamleh, D. B. Leinweber, \emph{arXiv: 1304.0325v2[hep-lat]} (2013).

\noindent [23] I. I. Kogan, A. Kovner, \emph{Phys. Rev. D} {\bf52(6)} 3719 (1995).

\noindent [24] T.A. DeGrand, R. D. Loft, \emph{Comp. Phys. Commu.} {\bf65}, 84 (1991).

\noindent [25] J.Dragos, R. Horsley, W. Kamleh, D. B. Leinweber, Y. Nakamura, P. E. L. Rakow, G. Schierholz, R. D. Young, J. M. Zanotti, \emph{Phys. Rev. D} {\bf94}, 074505 (2016).

\noindent [26] A. Vega, J. Flores, \emph{Pramana-J. Phys.} {\bf87(73)}, 1 (2016).

\noindent [27] Z. Ghalenovi, A. A. Rajabi, \emph{Acta. Phys. Pol. B} {\bf42}, 1849 (2011).

\noindent [28] H. M. Chot et al.,  \emph{Acta. Phys. Pol. B} {\bf6}, 281 (2013).

\noindent [29] NPTEL-Phase II, \emph{Advanced Statistical Physics}.

\vskip 5in

\end{document}